\documentclass[preprint]{aastex}		% LaTeX 2
\newcommand{\asinh}{\textrm{asinh}}
\newcommand{\includeG}[2]{%
  \includegraphics[width=#1]{#2.ps}}

\newcommand{\HDFdescription}{%
The top row of images were prepared using using
Eq. \ref{EqSeparateStretch}, while the bottom row used
Eq. \ref{EqIntensityStretch}.  The top left panel shows the familiar
STScI color image. The bottom left panel use an $\asinh$ stretch, and
is our preferred representation of the HDF. The top middle panel is
our best match to the STScI image and uses a logarithmic stretch; the
middle bottom panel uses the same logarithmic function (i.e. identical
values of $m$ and $M$). The righthand pair of panels share a hard
linear stretch (it is identical to the linear part of the $\asinh$
stretch of the bottom left panel).}%

\begin{document}
\title{Preparing Red-Green-Blue (RGB) Images from CCD Data}
\author{Robert Lupton\altaffilmark{1}}
\email{rhl@astro.princeton.edu}
\author{%
  Michael~R.~Blanton\altaffilmark{2},
  George Fekete\altaffilmark{3},
  David~W.~Hogg\altaffilmark{2},
  Wil O'Mullane\altaffilmark{3},
  Alex Szalay\altaffilmark{3},
  N.~ Wherry\altaffilmark{2}
  }
\altaffiltext{1}{Princeton University Observatory, Princeton, NJ 08544, U.S.A.}
\altaffiltext{2}{Center for Cosmology and Particle Physics, Department of Physics, New York University, NY 10003, U.S.A.}
\altaffiltext{3}{Johns Hopkins University, Baltimore, MD 21218, U.S.A.}

\begin{abstract}

We present a new, and we believe arguably correct, algorithm for
producing Red-Green-Blue (RBG) composites from 3-band astronomical
images. Our method ensures that an object with a specified
astronomical color (e.g. $g-r$ and $r-i$) has a unique color in the
RGB image, as opposed to the burnt-out white stars to which we are
accustomed. A natural consequence of this is that we can use the
\emph{same} colors to code color-magnitude diagrams, providing a
natural `index' to our images.

We also introduce the use of an $\asinh$ stretch, which allows us to
show faint objects while simultaneously preserving the structure of
brighter objects in the field, such as the spiral arms of large
galaxies.

We believe that, in addition to their aesthetic value, our images
convey far more information than do the traditional ones, and provide
examples from Sloan Digital Sky Survey (SDSS) imaging, the Hubble Deep
Field (HDF), and Chandra to support our claims. More examples are
available at
\texttt{http://www.astro.princeton.edu/$\sim$rhl/PrettyPictures}.

\end{abstract}

\keywords{techniques: image processing, techniques: photometric}

\maketitle

%=-=-=-=-=-=-=-=-=-=-=-=-=-=-=-=-=-=-=-=-=-=-=-=-=-=-=-=-=-=-=-=-=-=-=-=-=-=-=-

\section{Introduction}

\textit{%
  The purest and most thoughtful minds are those which love colour the most}
\footnote{John Ruskin, The Stones of Venice}

One of the reasons that our forbears were drawn to astronomy was the
sheer drama and beauty of the night sky.  When photography became a
standard astronomical technique, it was natural to combine images
taken in different bands to produce spectacular colored images of the
night sky (e.g. \cite{Malin} and references therein).

When CCD detectors were introduced into astronomy 25 years ago, and
especially when large-format cameras became common, astronomers were
once again motivated to generate 3-color composites.  CCDs are
approximately linear, unlike the roughly logarithmic sensitivity curve
of photographic plates, so the natural way to produce an image was to
map the logarithm of the pixel intensities to the red, green, and blue
brightnesses in the final image (see e.g. \cite{Villard}). This
algorithm produces pictures that look rather like those produced from
photographs, but unfortunately it also discards a lot of information.

\section{Generating Colored Images from CCD Data}

Let us consider CCD data taken in three bands which we shall call $b$,
$g$, and $r$ (E.g. $B$, $R$, and $I$); we assume that these data
have been background-subtracted, if necessary linearised, flatfielded,
and scaled appropriately. This scaling is to some extent arbitrary, as
different astronomers prefer different color balances, but we have
found that a conversion to $f_\nu$ (in practice to an AB scale,
\cite{AB}) works well.

We desire a mapping to the
range $[0,1]$ for each of three colors red (R), green (G), and blue
(B). The usual algorithm is:
\begin{eqnarray}
R = f(r);\, G = f(g);\, B = f(b)
\label{EqSeparateStretch}
\end{eqnarray}
where
$$
f(x) = \cases{0,     & $x < m$; \cr
              F(x - m)/F(M - m) & $m \le x \le M$; \cr
              1 & $M < x$.
	      }
$$ and $m$ is the minimum value to display, and $M$ the maximum; these
are chosen to bring out favoured details. As mentioned in the
introduction, a common choice is logarithmic ($F \equiv \ln$), but
linear or square-root stretches are also popular. The actual implementation,
at least for integer input data, often uses a lookup table.

This style of mapping simulates one of the unfortunate features of
photographic plates, namely that all pixels for which $r$, $g$, and $b$
are $\ge M$ appear white in the final image.  In a photograph this
white color corresponds to a lack of information, but for a CCD it
is a choice made by the astronomer.  In fact, for \emph{any} non-linear
function $F$, an object's color in the composite image depends upon
its brightness, even if \emph{none} of $r$, $g$, or $b$ exceed $M$.

Fortunately, we can easily avoid these problems.
Defining $I \equiv (r + g + b)/3$
\footnote{It is possible to use a different measure of $I$, e.g. a root-mean-square or $\sqrt{\chi^2}$ value.}, set
\begin{eqnarray}
\label{EqIntensityStretch}
\nonumber R &= r*f(I)/I\\
          G &= g*f(I)/I\\
\nonumber B &= b*f(I)/I
\end{eqnarray}
\newcommand{\rlineto}{\equiv}%
If $\max_{RGB} \equiv \max(R, G, B) > 1$, set $R \rlineto
R/\max_{RGB}$, $G \rlineto G/\max_{RGB}$, and $B \rlineto B/\max_{RGB}$;
if $I \equiv 0$, set $R = G = B = 0$.
In other words, the intensity is clipped at unity, but the \emph{color} is
correct.
\footnote{We note that it is often possible to include WCS information
in the header of the RBG image; file formats that allow this include
JPEG and TIF.}
Additionally, it is possible to choose a more flexible
functional form for $F$; following \cite{asinhMags}, we take $F(x)
\equiv \asinh(x/\beta)$, where the softening parameter $\beta$ is
chosen to bring out desired details --- it determines the point at
which we change from a linear to a logarithmic transformation
($\asinh(x) \sim x$ for $x \ll 1$; $\asinh(x) \sim \ln(2x)$ for $x \gg
1$)
\footnote{A convenient parameterisation is $f(x) = \asinh(\alpha Q (x-m))/Q$,
  which allows the user to first set $Q \rightarrow 0$ and choose the linear
  stretch $\alpha$, and then adjust $Q$ to bring out brighter features. In this
  case, $M \equiv m + \sinh(Q)/(\alpha Q)$.}.
The linear part is used to show faint features in the data,
while the logarithmic part enables us to simultaneously illustrate
structures such as spiral arms or the cores of star clusters that are
significantly brighter. As an additional trick, we find that sometimes
a $\sqrt{\asinh}$ stretch brings out the details of dust lanes in
spiral galaxies; here, as before, the $\asinh$ is used to control the
dynamic range.

\begin{figure}
  \includeG{6in}{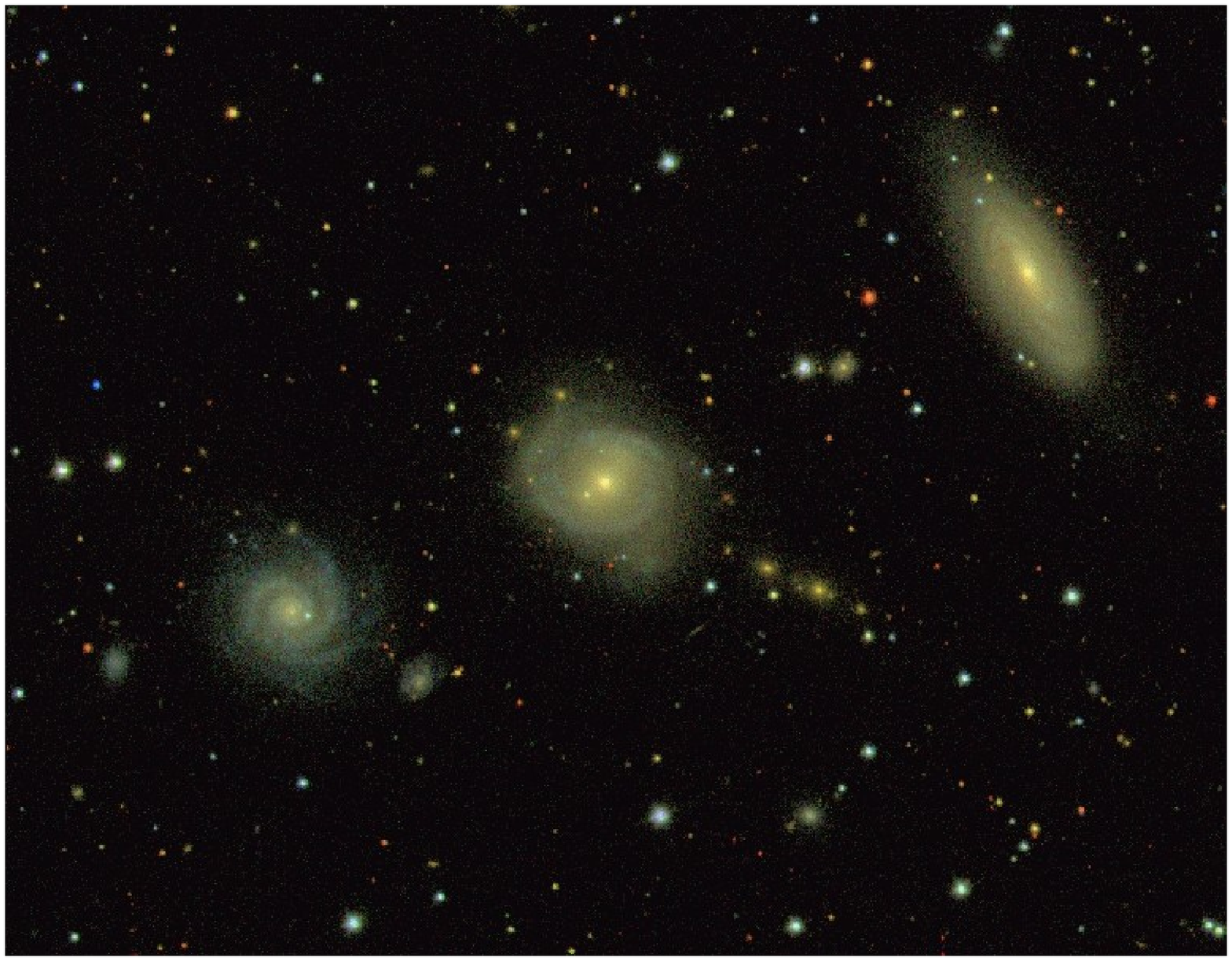}
\caption[NGC6978]{%
An approximately $5' \times 6.5'$ g-r-i composite based on SDSS imaging of
the galaxies NGC6976, NGC6977, and NGC6978. The faintest objects visible
in this image have $r \sim 22$. Note the colors of the stars, ranging
from bluish through yellow to red, the blue spiral disks, and the
yellowish color of the galaxies spheroids. The stretch is $\asinh$
with $Q \equiv 8$ and $\alpha \equiv 0.02$.
}
\label{figNGC6978}
\end{figure}

The results of this simple change are dramatic; a given color
(i.e. value of $g-b$ and $r-g$) is now mapped to a unique color in the
RGB image.  Fig.  \ref{figNGC6978} shows an SDSS \citep{York} g-r-i
composite of the galaxies NGC6976, NGC6977, and NGC6978.  The reader
will notice that the stars, rather than all being white, show a
variety of red, yellow, and bluish tints; that the spiral bulges (and
the group of ellipticals to the lower right of NGC6977) are a uniform
yellowish color, while the spiral disks are blue ($[OIII]$ and
$H_\alpha$ appear in B and R respectively). Furthermore, it is evident
that star formation activity in the disks (as indicated by the
distinctive bluish color) decreases from left to right.

\begin{figure}
  \includeG{6in}{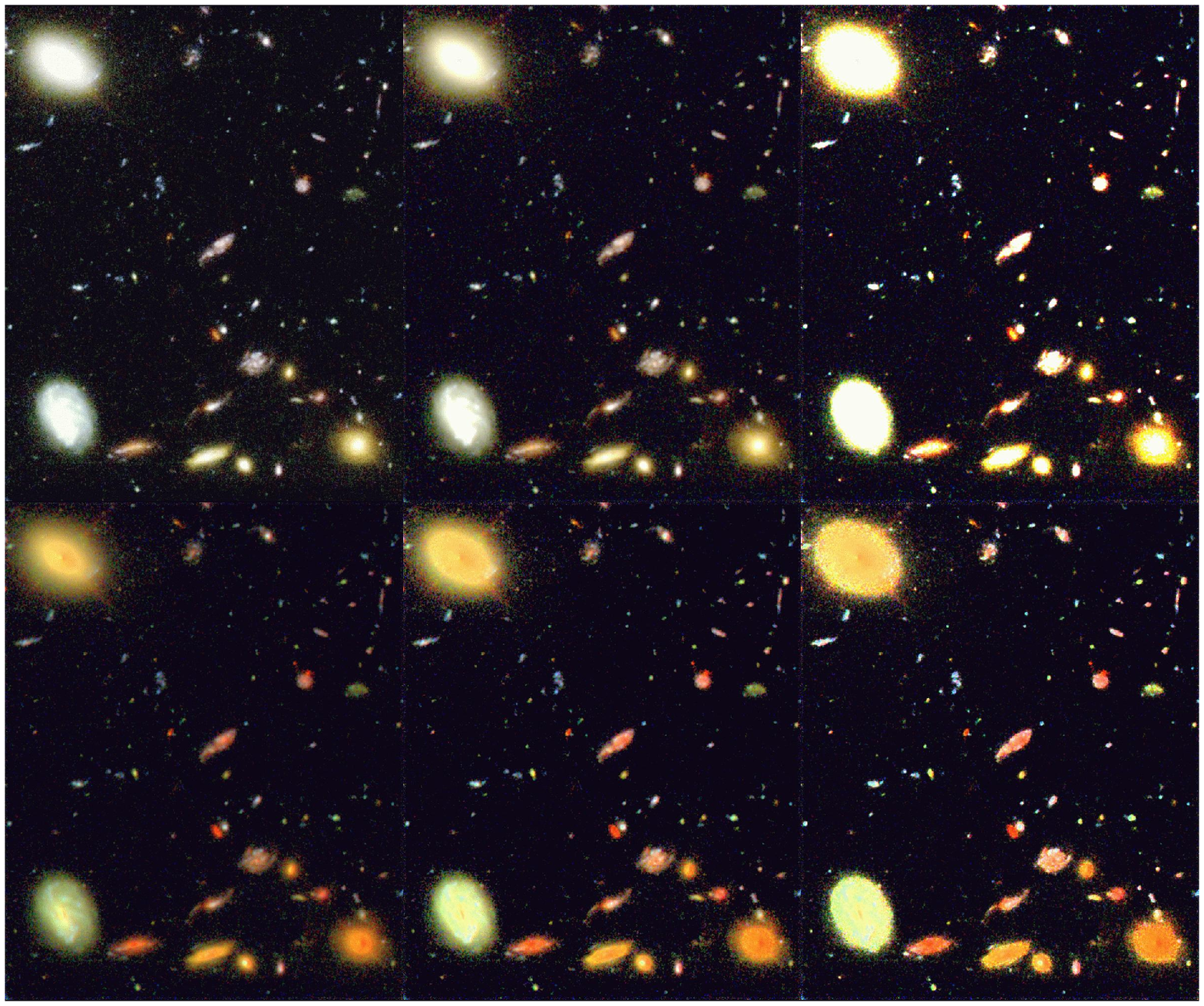}
\caption[A portion of the HDF]{%
A $f450$, $f606$, and $f814$ composite of the HDF
\footnote{Picture credit: NASA, R. Williams (STScI) and the Hubble
Deep Field Team.}.
\HDFdescription{} %
In the bottom row of images, the larger galaxies have similar colors
to the galaxies in Fig. \ref{figNGC6978}, but the fainter, more
distant, red galaxies have no analogues in the shallow SDSS data; our
eyes are seeing the effects of redshifting the 4000\AA{} break.
}
\label{figHDF}
\end{figure}

We may use the well-known color composite of the Hubble Deep Field
(HDF; \cite{HDF}), Fig. \ref{figHDF}, as a pedagogical
example.
\HDFdescription

Interestingly, the bottom set of panels are pretty similar, with the
main difference being that the spiral arms in the lower left corner are
rather better defined in the leftmost panel. It is clear that much
of the morphological information is carried in the \emph{color}
rather than the \emph{intensity}.

The bands ($f450$, $f606$, and $f814$) used in Fig. \ref{figHDF} are
not quite the same as those used in the SDSS composites ($g$, $r$, and
$i$) but the astrophysical palette is similar, in the sense that the
larger HDF galaxies have colors similar to those seen in
Fig. \ref{figNGC6978}, with blue disks and yellowish bulges. The
fainter objects, however, look unlike \emph{anything} seen in SDSS
imaging, with bright red (K-corrected) bulges and, in some cases,
bluish disks which must be due to active star formation and the
emission of far-uv light.

\begin{figure}
\includeG{6in}{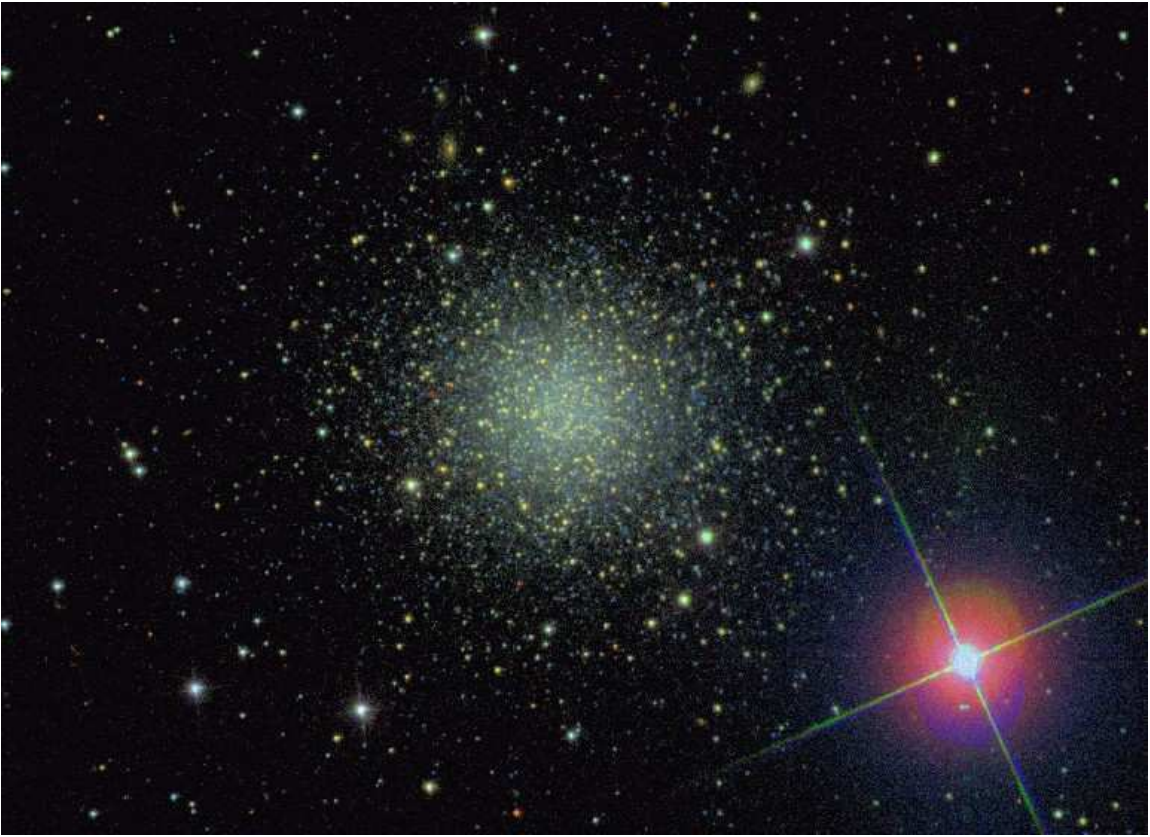}
\caption[NGC2419]{%
An approximately $7'\times 9'$ g-r-i composite based on SDSS imaging of
the halo globular cluster NGC2419. Note the blue horizontal branch and
yellowish red giants; the red stars are M-stars in the disk. The
diffraction spikes show that the instrument rotator moved
significantly during the 5 minutes between the $g$ and $r$ exposures.
The stretch is $\asinh$ with $Q \equiv 9$ and $\alpha \equiv 0.02$.
}
\label{figNGC2419}
\end{figure}

\begin{figure}
  \includeG{4in}{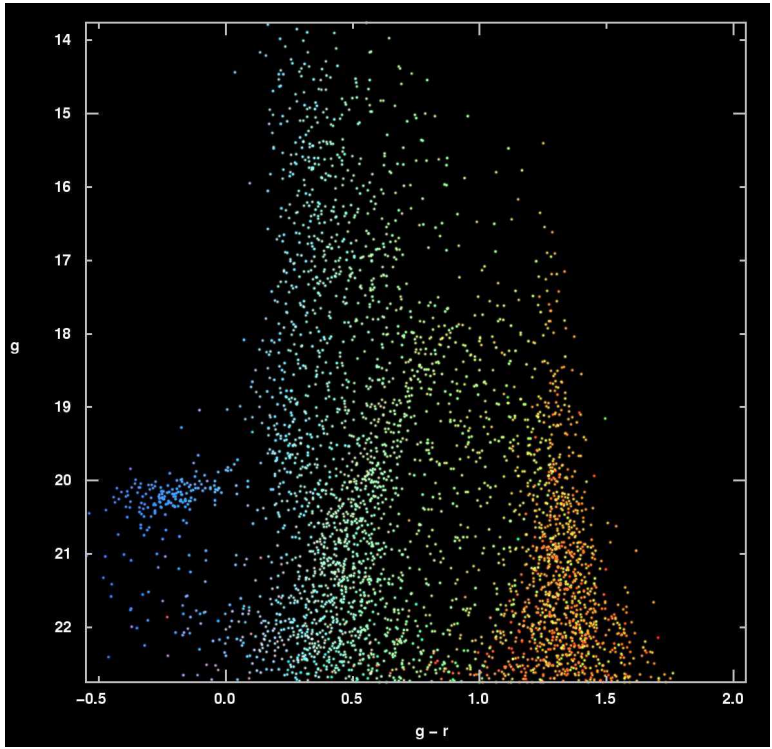}
  \caption[NGC2419-gr]{%
    A $g-r$ v. g color-magnitude diagram for stars in an area of about
    0.6 square degrees around the cluster NGC2419. Each star has the
    same color as it has is in a gri composite image of
    that part of the sky, a part of which is shown in Fig. \ref{figNGC2419}.
  }
  \label{figNGC2419-gr}
\end{figure}

Fig. \ref{figNGC2419} shows the globular cluster NGC2419;
Fig. \ref{figNGC2419-gr} shows a color-magnitude diagram of point
sources in a 0.6 square degree region surrounding the cluster; the
points in this diagram have the \emph{same} color as the corresponding
objects in the image; the cluster horizontal- and giant branches
stand out.

\begin{figure}
  \includeG{6in}{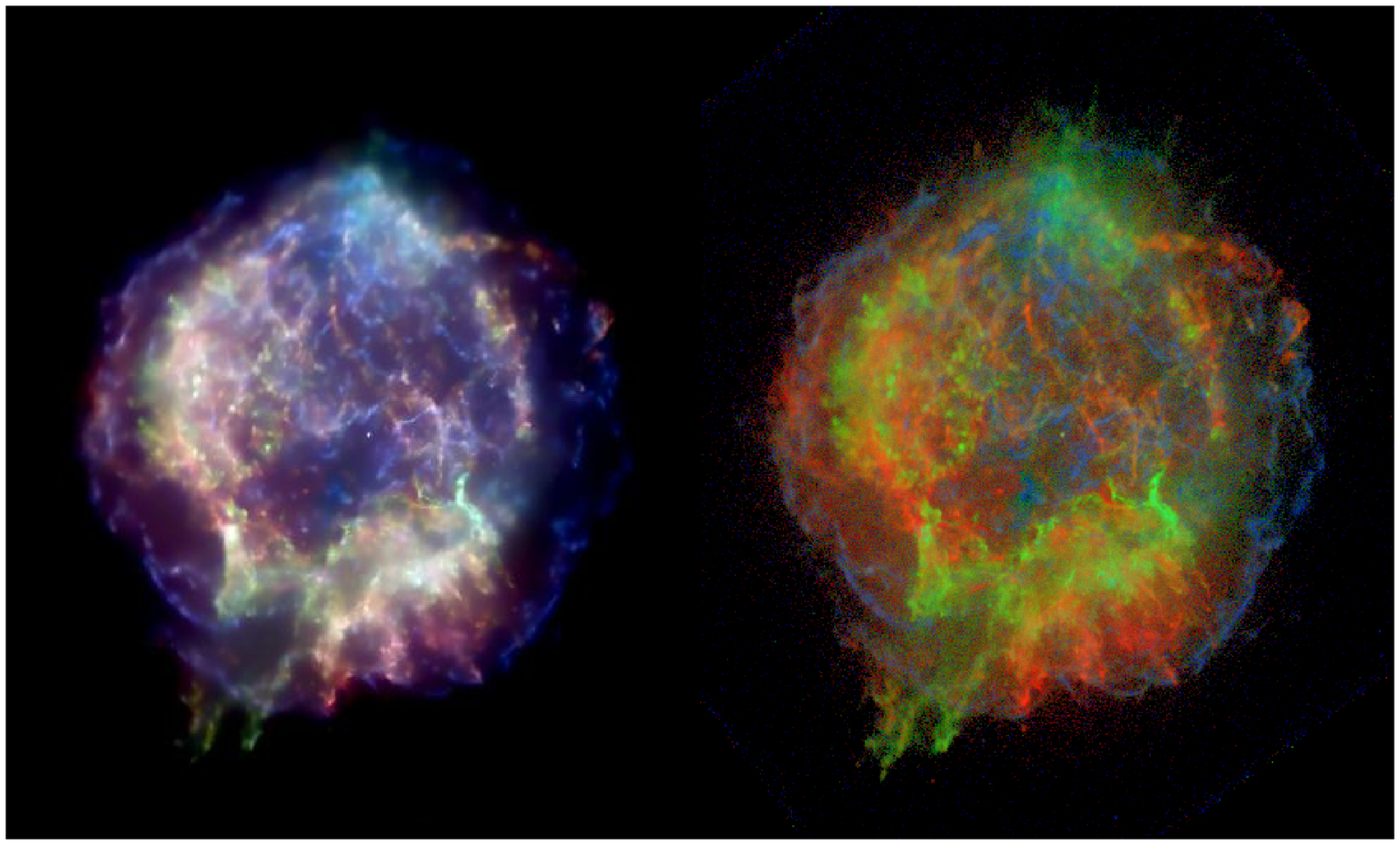}
  \caption[Chandra image of CasA]{%
    The Supernova remnant Cas A, as
    imaged by the Chandra X-ray satellite using the ASIS camera
    \footnote{Data and CXC color composite prepared by NASA/CXC/SAO}.
    R, G, and B correspond to 0.3---1.55 keV, 1.55---3.34 keV, and
    3.34---10keV respectively. The lefthand image is from the Chandra
    Supernova Catalog, the righthand image is prepared using
    Eq. \ref{EqIntensityStretch} and an $\asinh$ stretch. The input
    image is the raw Chandra data smoothed with a sigma=0.5pixel
    Gaussian; this is not the same as the inputs to the left hand panel.
    Note the presence of both blue and red filaments crossing the center
    of the remnant. It is possible to generate a figure analogous
    to Fig. \ref{figNGC2419-gr} that shows the mapping from hardness
    ratios to RGB.
  }
  \label{figCasA}
\end{figure}

Finally, this technique is not restricted to optical data; Fig.
\ref{figCasA} shows a Chandra X-ray image of the supernova remnant Cas A.

\section{Detector Saturation}
\label{saturation}

Although CCDs are wonderful detectors, they are not perfect. The SDSS
data processing pipeline \citep{photo} interpolates over saturation
trails, which accounts for their absence from these images.
Unfortunately, it is not able to correctly recover the counts in the
cores of saturated stars, but we \emph{do} know which pixels were
involved\footnote{If this information isn't available, all pixels
  above some threshold may be considered tainted.}.

Eq. \ref{EqIntensityStretch} means that every pixel in the image
preserves its color, even in the saturated cores of stars. We can
avoid undesirable artifacts in the final picture by finding connected
pixels which were saturated in at least one band, and replacing them
with the average color of the pixels touching the saturated region.
Because we do not wish to average over the
long bleed-trails produced by bright stars, we only apply this
procedure to saturated pixels with intensities $> M$.
An obvious example of a bright star is seen at the bottom right of
Fig. \ref{figNGC2419}; in this case we treated all pixels
with at least 10000 counts in any band as `saturated'.

\section{Practical Implementations}

The original implementation of these algorithms was included in the
SDSS image processing code, but fortunately stand-alone versions are
also available.  An IDL version is available from
\texttt{http://cosmo.nyu.edu/hogg/visualization/rgb/}, and a
C version from\hfil\break
\texttt{http://www.sdss.jhu.edu/doc/jpeg.html}.
\section{Conclusions}

Color images have traditionally been considered a luxury, but with the
techniques espoused in this paper, we have found that they convey an
enormous amount of information.  For example, star-forming regions
have a distinctive color, and it is very clear, looking at images of
spiral galaxies, how much the star formation rate varies. We encourage
the reader to visit\hfil\break~\qquad
\texttt{http://www.astro.princeton.edu/$\sim$rhl/PrettyPictures}
\hfil\break
where,
among other objects, all of the NGC galaxies in the SDSS DR1 \citep{DR1}
are presented.

\vskip1cm
RHL thanks Michael Strauss for helpful comments on an early version
of this manuscript.

Funding for the creation and distribution of the SDSS Archive has been
provided by the Alfred P. Sloan Foundation, the Participating
Institutions, the National Aeronautics and Space Administration, the
National Science Foundation, the U.S. Department of Energy, the
Japanese Monbukagakusho, and the Max Planck Society.  The SDSS Web
site is \texttt{http://www.sdss.org/}.

The SDSS is managed by the Astrophysical Research Consortium (ARC) for
the Participating Institutions.  The Participating Institutions are
The University of Chicago, Fermilab, the Institute for Advanced Study,
the Japan Participation Group, The Johns Hopkins University, Los
Alamos National Laboratory, the Max-Planck-Institute for Astronomy
(MPIA), the Max-Planck-Institute for Astrophysics (MPA), New Mexico
State University, University of Pittsburgh, Princeton University, the
United States Naval Observatory, and the University of Washington.

%=-=-=-=-=-=-=-=-=-=-=-=-=-=-=-=-=-=-=-=-=-=-=-=-=-=-=-=-=-=-=-=-=-=-=-=-=-=-=-

\vfil\eject
\section{Figures}

%----------------------------------------

%----------------------------------------

%----------------------------------------

\begin{thebibliography}{99}

\bibitem[Abazajian et al., 2003]{DR1}
Abazajian, K., et al., 2003, AJ 126, 2081.

\bibitem[Lupton et al., 1999]{asinhMags}
Lupton, R. H., Gunn, J. E. \& Szalay, A. 1999, AJ 118, 1406.

\bibitem[Lupton et al., 2001]{photo}
Lupton, R.~H. et al. 2001 ADASS X, 269.

\bibitem[Oke and Gunn, 1983]{AB}
Oke, J.B. and Gunn, J.E., 1983. ApJ 266, 713.

\bibitem[Malin, 1992]{Malin}
Malin,~D., 1992. Royal Astron. Soc. Quart. Jrn. 33, 321.

\bibitem[Villard and Levay, 2002]{Villard}
Villard, R, and Levay, Z 2002. Sky~\&~Telescope 104, V3, 28.

\bibitem[Williams et al, 1996]{HDF}
Williams, R.~E., et al. 1996. AJ 112, 1335.

\bibitem[York et al., 2000]{York}
York et al., 2000 AJ 120 1579.

\end{thebibliography}
\end{document}